\documentclass{aastex63}
\submitjournal{ApJL}
\turnoffediting
\begin{document}
\title{\edit2{A possible transit of a disintegrating exoplanet in the nearby multiplanet system DMPP-1}}
%\title{Detection thresholds and a possible transit of a catastrophically disintegrating exoplanet in TESS photometry of the nearby, ablating, compact multiplanet system DMPP-1}
%\title{Constraints on the nature of the nearby ablating planetary system DMPP-1 from TESS photometry}

\correspondingauthor{M.H.Jones}
\email{m.h.jones@open.ac.uk}

\author[0000-0001-6217-8574]{Mark H. Jones}
\affiliation{School of Physical Sciences, The Open University, Milton Keynes, MK7~6AA, U.K.}

\author[0000-0002-8050-1897]{Carole A. Haswell}
\affiliation{School of Physical Sciences, The Open University, Milton Keynes, MK7~6AA, U.K.}

\author[0000-0001-6105-2902]{John R. Barnes}
\affiliation{School of Physical Sciences, The Open University, Milton Keynes, MK7~6AA, U.K.}

\author{Daniel Staab}
\altaffiliation{AVS, Rutherford Appleton Laboratory, Harwell, Oxford, U.K.}
\affiliation{School of Physical Sciences, The Open University, Milton Keynes, MK7~6AA, U.K.}

\author[0000-0002-9831-0984]{Ren\'{e} Heller}
\affiliation{Max Planck Institute for Solar System Research, Justus-von-Liebig-Weg 3, 37077 Göttingen, Germany}

\received{2020 March 17}
\revised{2020 April 30}
\accepted{2020 May 1}
% revision history
% 13 March 2020 CH radical edits for conciseness, new v4
% 20 Feb 2020 CH add rocky planet mineral atmosphere models and abstract
% 20 Feb MHJ revise table and check numbers and consistency
% 18 Feb 2020 CH edits for interpretation, consistency and style
% 12 Feb 2020 CH edits for Fe planet limits and interpretation
% 10 Feb 2020 CH edits for CDE context
% 07 Feb 2020 CH edits after checking TESS publications
% 30 Jan 2020 MHJ Revised draft complete
% 23 Jan 2020 MHJ added discussion
% 23 Jan 2020 CH add intro
%
% n.b. ApJLett instructions:
% A Letter must also be concise and to the point. Authors should strive to prepare manuscripts within the following limits:

% Abstract – no more than 250 words
% Main Text – no more than 3500 words (not including appendices or other supplementary material)
% Figures and Tables – no more than 5 combined figures and tables, e.g. 3 figures and 2 tables.
% References – no more than 50 references

%
%\onehalfspacing

%\maketitle

\begin{abstract}
We analyse TESS photometry of DMPP-1 \edit1{(HD\,38677; TIC\,66560666), a nearby F8V star hosting hot super-Earth planets and a warm Neptune.} Using the Transit Least Squares algorithm \edit1{and other methods} we find a transit signal at 
$P=3.2854^{+0.0032}_{-0.0025}$\,d
with depth 87$^{+25}_{-30}$\,ppm \edit1{and false alarm probability 1.6\%. This is shallower than hitherto published TESS discoveries.}
The \edit1{3.285\,d} signal is recovered for several, but not all, \edit1{methods for detrending} stellar astrophysical variability. \edit1{Further observations are needed to improve the significance of the detection.} If this transit were due to an Earth-like rocky planet it would have been detected in the RV data, but it is not. The TESS data cover seven individual transits, one of which is consistent with zero depth. The insolation of the putative planet is  $990 {\rm S_{\oplus}}$, \edit1{typical} of fluxes experienced by the three known catastrophically disintegrating exoplanets (CDEs). The transits can be self-consistently attributed to a CDE with a mass below the RV detection threshold.  We searched for transits of the known RV planets, finding null results and detection thresholds of $< 100$ ppm, which we quantify for each. \edit1{The DMPP-1 planetary system was discovered as a consequence of circumstellar gas attributed to ablation of hot planets. The RV planets} may have been ablated to near-pure iron cores. We place limits on the orbital inclinations of the RV planets where the expected transit depth exceeds the detection threshold.  If the \edit1{3.2854 day} transit detection is confirmed, e.g. with CHEOPS photometry, DMPP-1 would be a first-rate target for JWST spectroscopy. %[249 words].

\end{abstract}

\section{Introduction}
% DMPP-1
% RV planets
% TESS 
%The first known transiting 
Terrestrial exoplanet research began with the discovery of CoRoT-7\,b
%, was discovered through the detection of its transits by the pioneering CoRoT space telescope 
\citep{2009A&A...506..287L}. Since then Kepler 
%has discovered thousands of transiting planets and revolutionised our knowledge of the Galaxy's planet content. Kepler 
has shown that small, presumably rocky planets are common
%, 
%[citation needed]
%and that compact multi-planet systems with several terrestrial-sized planets 
in orbits smaller than that of Mercury.
%are plentiful [{\bf citation needed}]. 
Among Kepler's most exciting rocky planet discoveries are the catastrophically disintegrating exoplanets (CDEs), with prototype Kepler-1520\,b \citep{2012ApJ...752....1R}. These
%CDEs are 
hot, low mass, rocky planets are
%thought to be 
heated to temperatures 
$\sim 2000$\,K at the substellar point so that the rocky surface sublimes. Dust 
%then 
condenses from the metal-rich vapour producing
%and the Kepler CDEs were detected through the 
variable-depth transits of the dust clouds orbiting with the undetectably small ablating planets.

%Kepler's observations were deep and narrow-field, so the typical Kepler exoplanet host star is faint. 
Kepler-1520 is a $V=16$ K star; 
%and is too faint to permit precise, high-resolution spectroscopic characterisation. B
brighter, nearby analogues and progenitors
%of Kepler-1520\, b 
would allow exciting opportunities to probe the mineral composition of rocky planets outside our own Solar System \citep{2018AJ....156..173B}. TESS \citep{2015JATIS...1a4003R} is 
%currently 
performing an (almost) all sky transit survey and should 
%be able to 
find such objects.

%Meanwhile, the radial velocity (RV) technique has discovered almost a thousand exoplanets through targeted high resolution spectroscopy of bright FGK and M stars. 
%{\bf MHJ: needs footnote}
% Which database used for this? Also, exoplanet.eu and NASA exoplanet archive differ  
The Dispersed Matter Planet Project \citep[DMPP,][]{2019NatAs.tmp....2H} is an RV planet search which was partly motiviated by the goal of detecting nearby CDEs and their progenitors. 
\edit1{WASP-12 and other host stars of close-orbiting planets are shrouded in gas which absorbs the stellar chromospheric emission \citep{Haswell2012,2013ApJ...766L..20F,2017MNRAS.466..738S}. Thus anomalously low chromospheric emission can indicate hot, mass-losing planets. DMPP uses the widely-adopted metric of stellar activity, $\log{R'_{\rm HK}}$, which is derived from the Mount Wilson S-index \citep{1984ApJ...279..763N}, to select bright FGK V stars with apparently anomalously low chromospheric emission. This is attributed to absorption by metal-rich circumstellar gas ablated from hitherto undiscovered hot planets. Because these are the easiest stars for RV planet discoveries, the hypothesis underlying DMPP was that many of the targets will host ablating low mass, presumably rocky, planets. CDEs constitute the most dramatic examples of such planets.}
%The DMPP targets are all nearby, bright stars so the putative undiscovered hot planets were expected to be low mass, as giant planets orbiting these stars are among the easiest exoplanets to find. Hence DMPP executes high cadence, high precision RV measurements designed to detect low mass planets with orbital periods of a few days or less.   
DMPP-1 (HD~38677) is a nearby bright analogue of the Kepler compact multiplanet systems \citep{2019arXiv191210792S} orbiting a F8V host star with $V=7.98$. The system contains at least three super-Earth planets ($M_{\mbox{\scriptsize{P}}} \sin i$ values of  1.8--9.6  $\rm{M_{\oplus}}$) 
interior to a Neptune-like planet on a 19~d orbit. 
%Angular momentum considerations suggest the g
Gas ablated from an orbiting planet will
%should 
be concentrated in the orbital plane, so 
%Consequently 
DMPP should preferentially pick out ablating planet systems with an edge-on orientation. Thus we expect higher than random transit probabilities for the DMPP systems. 
%DMPP-1 is the first of these to be observed by TESS.
This paper reports an analysis of the TESS light curve for DMPP-1.

In Section~\ref{datameth} we describe the TESS data and the removal of systematic instrumental and astrophysical trends; the details of the approach affect the results of the transit searches we describe in Section~\ref{trans}. Section~\ref{disc} discusses our findings and their broader implications. Finally, Section~\ref{conc} briefly summarises our conclusions.

\section{Data and methods}\label{datameth} 
% Details of the Sector 6 observation
% The light curve
% Removal of the variability, GP vs. S-G

\begin{figure}[htp]
\centering
\includegraphics[width=15cm]{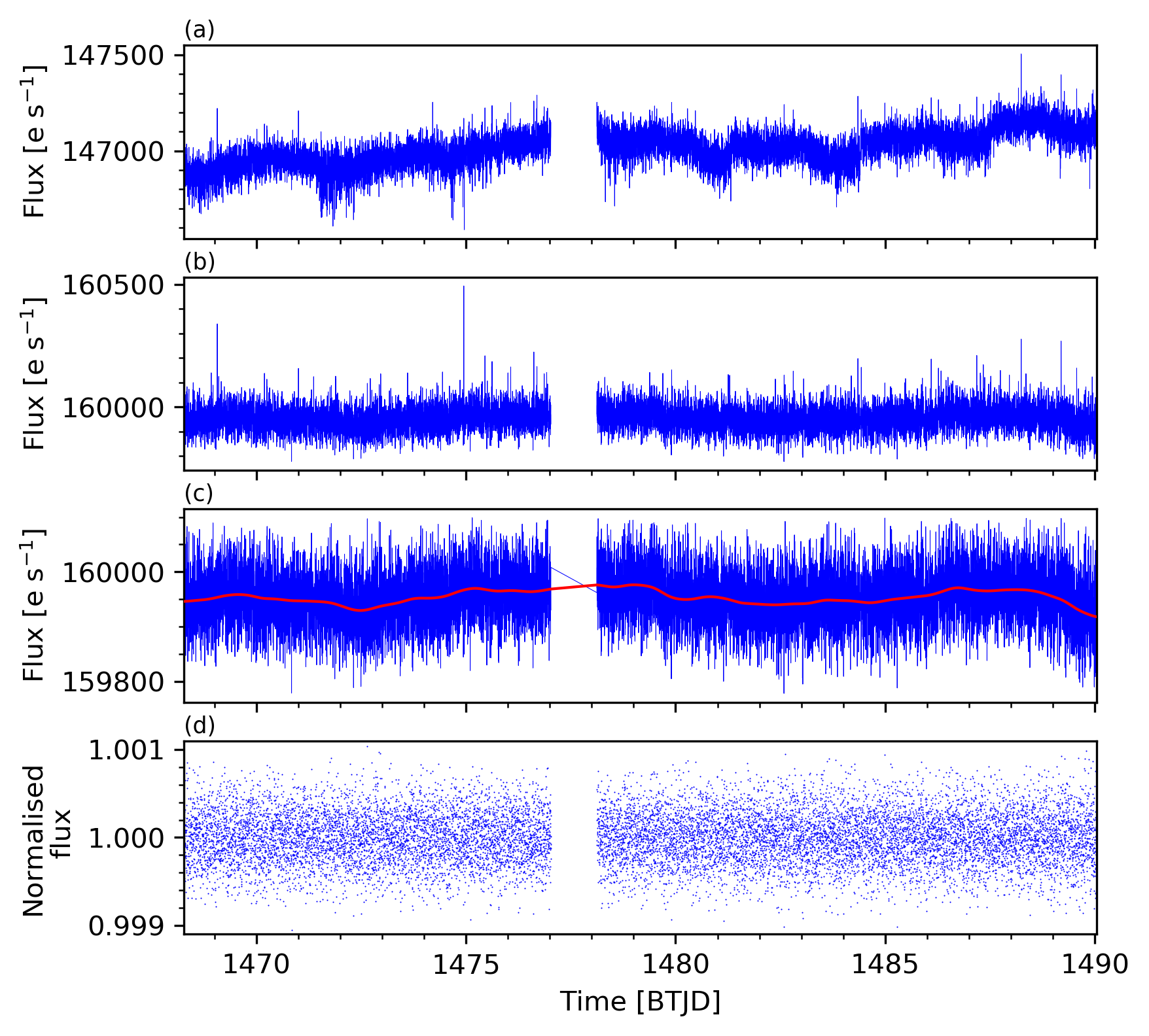}
\caption{(a) TESS Sector 6 SAP flux  of DMPP-1, (b) the corresponding PDC light curve. (c) The PDC flux light curve after sigma-clipping (blue) and the mean GP function (red).
%see text for details). 
(d) The cleaned, de-trended and normalised PDC flux used for transit searches.}
\label{fig:TESS_fluxes}
\end{figure}

TESS continuously covers 
%operates in a survey mode that observes an area of 
2304 square degrees (a Sector) for
%over two consecutive spacecraft orbits (
$\sim 27$ days
%) 
obtaining photometry \edit1{in the 600--1000\,nm bandpass} at 2 minute cadence on selected targets. DMPP-1 was observed with TESS Camera~2, CCD~4 during Sector 6 (December 2018); i.e., 
%The light-curve covers 
BTJD 1468.27 to 1490.04 (Barycentric Julian Date$-2457000.0$) with a $\sim 1.1$~d data gap at BTJD 1477.03. Pipeline processing 
%of TESS data 
produces a background-subtracted calibrated standard aperture photometry (SAP) light curve (Figure \ref{fig:TESS_fluxes}a). Subsequent removal of systematic instrumental effects 
%using co-trending basis vectors (CBVs) which model the variability common to all targets on a given camera/CCD combination follows, 
yields
%ing 
the Pre-search Data Conditioning (PDC) flux \citep[Figure \ref{fig:TESS_fluxes}b;][]{TM-2018-220036,2017ksci.rept....8S}
%The
%Much of the variability in the SAP flux of DMPP-1 (Figure \ref{fig:TESS_fluxes}a) is removed by co-trending, resulting in a 
%PDC light-curve (Figure \ref{fig:TESS_fluxes}b)
%has a 
%, with 
with a 
mean value of $1.5995 \times 10^{5}$~e~s$^{-1}$. The smooth residual modulation on a time-scale of several days is assumed to be astrophysical. 

%To prepare the PDC light curve for 
Before our transit searches we: (i) rejected data with quality flag values $<512$ \citep[][Table~28]{TM-2018-220036} leaving 14827 cadences; (ii) recursively removed 
%positive 
outliers exceeding the median flux (of the entire light curve) by more than three standard deviations (`sigma-clipping'),
%above the median, 
rejecting 69 data points (c.f. 20 points if the data were Gaussian distributed); 
(iii) detrended 
this cleaned
%the PDC 
light curve using 
%a 
Gaussian Processes (GP) 
%method 
as described below. Sigma-clipping was implemented using {\tt lightkurve} \citep{2018ascl.soft12013L}.

We modelled the smooth modulation with 
%using a GP approach using the 
{\tt Celerite} 
%software of 
\citep{2017AJ....154..220F}.
%. 
A GP kernel comprising a single damped harmonic oscillator with damping parameter $Q=1/\sqrt{2}$ is sufficient to model the smooth variability, finding a characteristic frequency $\omega_{0}=1.48$~rad~day$^{-1}$.
%\citep[see][Equation 24]{2017AJ....154..220F} 

%was found to be 

%The mean GP function is shown (in red) in Figure~\ref{fig:TESS_fluxes}c. 
Dividing by the mean GP function (red trace in Figure \ref{fig:TESS_fluxes}c)
yields the
%The cleaned data were divided by this function yielding the 
detrended, normalised light curve shown in Figure~\ref{fig:TESS_fluxes}d and used for subsequent analyses. GP detrending reduced the standard deviation of the sigma-clipped PDC light curve from 302~ppm to 293~ppm.

\section{Transit searches}\label{trans}
The RV measurements of DMPP-1 \edit1{were used to derive a preferred solution of four planets (see Table~\ref{tab:DMPP-1_RV_planets}a), and an alternative, less probable solution in which the data are modelled with an  S-index correlation term \citep[][Table~\ref{tab:DMPP-1_RV_planets}b]{2019arXiv191210792S}.}
%The solution without S-index correlation is preferred, though 
For completeness both solutions are considered 
here. 
%and summarised in Table~\ref{tab:DMPP-1_RV_planets}.  
The 
%TESS 
light curve spans 
%covers an interval of 
21.8~d, so 
%periodicity is detectable only for $ P < 10.9\, {\rm d}$. 
we searched for repeated transits at $ P < 10.9\, {\rm d}$ and for single and partial transits due to longer period planets. For clarity, we denote {\em expected} transit depths by $\delta$ and {\em measured} flux changes (or limits) by $(\Delta F/ F)$ (with measurement time-scale).

\begin{table}[htp]
\small
    \centering
    \begin{tabular}{lccccc} \hline
    \multicolumn{6}{c}{(a) Preferred solution} \\ \hline
                & DMPP-1 b & DMPP-1 c  & DMPP-1 d & DMPP-1 e &   \\
    $P$ [d]     & 18.57   & 6.584  & 2.882   & 5.516  &\\
    $a$ [AU]    & 0.1462  & 0.0733 & 0.0422  & 0.0651 & \\
    $M_{\mbox{\scriptsize P}} \sin i$ [M$_\oplus$] 
                & 24.27   & 9.60   & 3.35    & 4.13 & \\
 $S$ [${\rm S_{\oplus}}$]   &98 & 390 & 1177 & 494 & \\ 

$R_{\rm P, Earth-like} \,[{\rm R_{\oplus}}]$ &  & 1.814 & 1.393 & 1.471 & \\
$R_{\rm P, Fe} \,[{\rm R_{\oplus}}] $ &  &1.434 & 1.126&1.184 & \\
$R_{\rm P, H_{2}O} \,[{\rm R_{\oplus}}] $ &3.243 &   &  &  & \\
%$\delta$ (Fe, Earth-like, H$_{2}$O) [ppm] & -, -, 649 & 127, 203, - & 78, 120, - & 87, 134, - & \\
$\delta_{0}$ [ppm] & 649 & $203^{\dag}, 127^{\ddag}$ & $120^{\dag},78^{\ddag}$  & $134^{\dag},87^{\ddag}$ & \\
 Method          & `step'  & `fold' & `fold'  & `fold' \\
    $T_{\mbox{\scriptsize{dur, max}}}$ [d]
                & 0.237   & 0.168  & 0.127   & 0.158  \\
$(\Delta F / F)_{\mbox{\scriptsize{max}}}$ [ppm]  
                & $<158$ & $<93$ & $<57$ & $<78$ & \\
$b$ (Earth-like) & $>0.974^{\sharp}$ & $ >0.880$ & $ >0.877$ & $ >0.856$ & \\
$b$ (pure iron) &  & $>0.591$ & $>0.621$ & --- & \\ 
$i$ (Earth-like)  & $<87.76^{\circ \sharp}$ & $<85.96^{\circ}$ & $<83.00^{\circ}$ & $<85.93^{\circ}$ & \\
  $i$ (pure iron)   &  & $<87.26^{\circ }$ & $<85.05^{\circ}$ & --- & \\
%    & & & & & \\  
 \hline
\multicolumn{6}{c}{(b) Solution with S-index correlation} \\ \hline
                & DMPP-1 b$'$& DMPP-1 c$'$  & DMPP-1 d$'$ & DMPP-1 e$'$ & DMPP-1 f$'$  \\
    $P$ [d]     & 19.61   & 34.74      & 3.144   & 6.454   & 2.281   \\
    $a$ [AU]    & 0.1517  & 0.2220 & 0.0447  & 0.0723 & 0.0361 \\
    $M_{\mbox{\scriptsize P}} \sin i$ [M$_\oplus$] 
                & 19.92   & 20.12   & 3.95    & 4.95   & 1.82 \\
              
  $R_{\rm P, Earth-like} \,[{\rm R_{\oplus}}]$ &  &  & 1.453 & 1.540 & 1.181 \\        
   $R_{\rm P, Fe} \,[{\rm R_{\oplus}}]$ &  &  &1.172 &1.235 &0.965 \\  
     $R_{\rm P, H_{2}O} \,[{\rm R_{\oplus}}] $ &3.120 & 3.215  &  &  &  \\            
$\delta_{0}$ [ppm] &  601 & 603 & $130^{\dag},85^{\ddag}$ & $146^{\dag},94^{\ddag}$ & $86^{\dag},57^{\ddag}$ \\
Method          & `step'  & `step' & `fold'  & `fold' & `fold' \\
$T_{\mbox{\scriptsize{dur, max}}}$ [d]
                & 0.241   & 0.292  & 0.131   & 0.167  & 0.118  \\
 $(\Delta F / F)_{\mbox{\scriptsize{max}}}$ [ppm]  
                & $<156$ & $<142$ & $<65$ & $<95$ & $<55$\\
$b$ (Earth-like) & $>0.970^{\sharp}$ & $>0.976^{\sharp} $ & $>0.856$ & $>0.699$ & $>0.738$  \\
$b$ (pure iron) &  &  & $>0.531$ & --- & ---  \\
$i$ (Earth-like)            & $<87.85^{\circ \sharp}$ & $<88.52^{\circ \sharp}$ & $<83.55^{\circ}$ & $<86.75^{\circ}$ & $<83.11^{\circ}$ \\
$i$ (pure iron)        &  &  & $<86.01^{\circ}$ & --- & --- \\
 \hline
\multicolumn{6}{c}{(c) Marginal detection} \\ \hline
$R_{\mbox{\scriptsize{P}}}$ [$R_{\oplus}$]  & $1.38^{+0.29}_{-0.32}$  & & & \\  
$P$ [d] & $3.2854^{+0.0032}_{-0.0025}$ & & & & \\
$T_{0}$ [BTJD] & $1469.982\pm{0.010}$ & & & & \\
$i$      & $ {83.4^{\circ}}^{+1.0^{\circ}}_{-0.4^{\circ}} $  & & & & \\
$a$ [AU] & $0.0461\pm{0.0008}$ & & & & \\
$T_{\mbox{\scriptsize{dur}}}$ [d] & $0.06^{+0.03}_{-0.01}$ & & & & \\
$b$ & $0.90^{+0.04}_{-0.13}$ & & & & \\
$\delta$ [ppm] & $87^{+25}_{-30}$ & & & & \\
$M_{\mbox{\scriptsize{P, Earth-like}}}$ [M$_\oplus$] & $3.2^{+3.6}_{-2.0}$ & & & \\

%$(\Delta F / F)$ [ppm] & 95 & & & & \\
%$T_{\mbox{\scriptsize{dur}}}$ [d] & 0.06264 & & & & \\
%$M_{\mbox{\scriptsize{P, Fe}}}$ [M$_\oplus$]  & 9.57 & & & & \\
    \end{tabular}
    \normalsize
    \caption{ (a), (b)  DMPP-1 RV planets from \cite{2019arXiv191210792S} with radii, $R_{\mbox{\scriptsize{P}}}$, using the Earth-like, pure iron (Fe) and 100~percent water at $T=1000$~K (H$_{2}$O) models of \cite{Zeng19}. $\delta_{0}$ is the expected transit depth when impact parameter $b=0$ ($\dag$ and $\ddag$ indicate Earth-like and pure iron models respectively). `fold' and `step' denote the search methods. 
    %in Sections~\ref{sec:fold} and \ref{sec:step}. 
    For $b$ and $i$, $\sharp$ indicates the  100~percent water model ($T=1000$~K), and entries marked `---' are below the detection limit.  
    (c) The marginally detected signal and implied planetary parameters.  
    }
    \label{tab:DMPP-1_RV_planets}
\end{table}

\subsection{Signal limits at given periods}\label{sec:fold}
%Six of t
The Keplerian \edit1{RV} modulations of DMPP-1\,c, d, e, d$^{\prime}$, e$^{\prime}$ and f$^{\prime}$ 
%listed in 
(Table~\ref{tab:DMPP-1_RV_planets}) have 
%orbital period 
$P < 10.9$~d, 
so we folded
%allowing 
the light curve 
%to be folded 
to search for modulation 
using
%. The folded light curve was binned with 
a range of phase binning. 
%Our binning strategy maximises the statistical significance while retaining sensitivity to short transits and is robust to the exact timing of the transit with respect to that of the first data point.
The maximum bin width used was $(T_{\mbox{\scriptsize{dur, max}}}/P)=R_{\ast}/(\pi a)$,  where $T_{\mbox{\scriptsize{dur, max}}}$ is the maximum transit duration (i.e. for inclination $i=90^{\circ}$) of a planet in a circular orbit, $a$ is the semi-major axis, and $R_{\ast}$ is the stellar radius. The minimum bin width was 0.05 of the maximum value, and ranges from 8.48~minutes (DMPP-1~f$'$) to 12.07~minutes (DMPP-1~c). Every binning searched 10 starting ephemerides spread uniformly over a bin width.
%[{\bf MHJ: please check numbers. Checked and amended}]. 
The significance of any decrease in normalised flux from the mean value of 1.00 is expressed in terms of the standard error of the mean (derived from individual measurement uncertainties) in the corresponding phase bin ($\sigma_{\mbox{\scriptsize bin}}$). 
%in the normalised flux. 
For every folding, the most significant decrease (`dip') was identified. 

\begin{figure}[htp]
    \centering
    \includegraphics[width=15cm]{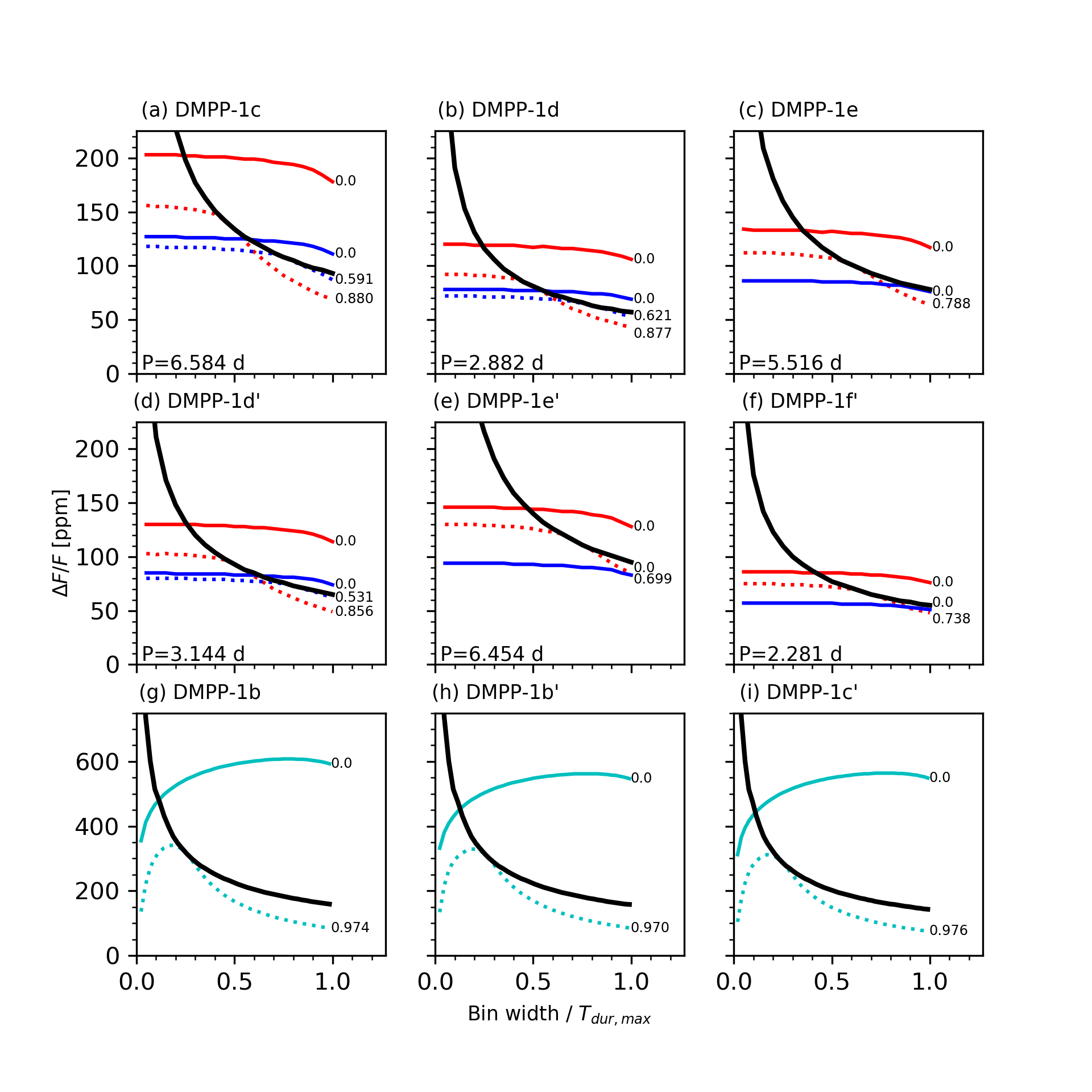}
    \caption{The $5 \sigma$ limits (black line) on the amplitude of transit signals from DMPP-1 planets \citep{2019arXiv191210792S} as a function of measurement time-scale (expressed as a fraction of the \edit1{maximum transit duration, $T_{\mbox{\scriptsize{dur, max}}}=P R_{\ast}/\pi a$)}. Coloured lines show the responses of the search methods with \cite{Zeng19} density models: red -- Earth-like; blue -- pure iron; cyan -- 100~percent water at 1000~K. Solid lines show the expected response when $b=0$. Dotted lines show the response at the limit of detectability along with the corresponding value of $b$. Panels (a)--(f) are calculated using the period folding method while panels (g)--(i) are based on the step-detection method (see text for details). }
    \label{fig:dip_step_limits}
\end{figure}

We found no signal with a significance exceeding $ 5\sigma_{\mbox{\scriptsize bin}}$. The  5$\sigma_{\mbox{\scriptsize bin}}$ upper limits (at a given bin width) are shown as solid black lines in Figure~\ref{fig:dip_step_limits}a--f, with the limit at maximum bin size $(\Delta F / F)_{\mbox{\scriptsize{max}}}$ listed in  Table~\ref{tab:DMPP-1_RV_planets}. 

To assess expected transit signals we created synthetic light curves using the {\tt batman} package \citep{2015PASP..127.1161K}, analysing them as above.  Planetary radii, $R_{\rm P, Earth-like}$, are estimated from the minimum masses $M_{\rm P}\sin{i}$ given by \cite{2019arXiv191210792S}, assuming $i=90^{\circ}$ and the Earth-like rocky planet models of \cite{Zeng19}. Semi-major axes of (assumed circular) planetary orbits are as given by \cite{2019arXiv191210792S}. 
We adopt $R_{\ast}=1.26 R_{\odot}$ \citep{2019arXiv191210792S} with a quadratic limb-darkening law ($u_{q}=0.3216$, $v_{q}=0.2301$) appropriate for a star with $T_{\mbox{\scriptsize{eff}}}=6200$~K and $\log g = 4.5$ \citep{2018AJ....156..102S}.  
 
\edit1{The responses of the search method to synthetic light curves for models with} impact parameter $b=0$ are shown (solid red lines) in Figures~\ref{fig:dip_step_limits}a--f noting that they approach the expected analytic transit depths $\delta_{0}$ \citep[][ see Table~\ref{tab:DMPP-1_RV_planets}]{2019A&A...623A.137H} as bin widths become small. All signals (for $b=0$, $i=90^\circ$) exceed the $5\sigma_{\mbox{\scriptsize bin}}$ detection limits and are ruled out by the data.

We investigated the detectability of grazing transits by varying $b$.  The dotted red lines in Figure~\ref{fig:dip_step_limits}a--f show the minimum value of $b$ for which the \edit1{ response to the model} lies below 
%the 
$ 5\sigma_{\mbox{\scriptsize bin}}$ at all bin widths. These limits on $b$ (and corresponding limits on $i$) are reported in Table~\ref{tab:DMPP-1_RV_planets}. 

%% MHJ - is there a shorter way to express this?
%The planets in DMPP-1 are among the most irradiated known RV planets, and lie along the lower boundary of the Neptune desert \citep{2019arXiv191210792S}. The system was orginally targeted by DMPP due to evidence for a circumstellar gas shroud attributed to planetary ablation \citep{2019NatAs.tmp....2H}. Thus we also considered the limiting case of planets which have been ablated down to an iron core: this produces the smallest conceivable planet radii for the measured masses. We used the pure iron planet models of \cite{Zeng19}
% the bibcodes provide unique labels, but the are utterly unmemorable. I find it much easier to label papers like this so I don't have to search for the label and cut and paste.
%to derive the radii corresponding to the minimum masses of these planets (see Table~\ref{tab:DMPP-1_RV_planets}). We establish that if planets have been ablated down to iron cores, DMPP-1c, -1d and 1d$^{\prime}$ are potentially detectable, but only over a reduced range of impact parameter ($b<0.591,0.621$ and $0.531$ respectively), while DMPP-1e, -1e$^{\prime}$ and -1f$^{\prime}$ would be too small to produce detectable transits. 

%We also considered the limiting case of planets which have been ablated down to an iron core: this produces the smallest conceivable planet radii for the measured masses. 

We also considered the case of planets which are predominantly iron (blue lines in Fig.~\ref{fig:dip_step_limits}a--f): this produces the smallest conceivable planet radii for the measured masses. As discussed by \cite{PriceRogers2019} iron planets may arise from collisionally-driven mantle stripping or formation from iron-enriched material, but as discussed in Section~\ref{sec:insolation}, we propose that mantle ablation is a key process in DMPP-1. 
We used the pure iron planet models of \cite{Zeng19}
to derive the radii corresponding to the minimum masses of these planets (see Table~\ref{tab:DMPP-1_RV_planets}). We establish that if planets have been ablated down to iron cores, DMPP-1~c, d and d$^{\prime}$ are potentially detectable, but only over a reduced range of impact parameters ($b<0.591,0.621$ and $0.531$ respectively), while DMPP-1~e, e$^{\prime}$ and f$^{\prime}$ would be too small to produce detectable transits. 

\subsection{Searches for Single Transits and Partial Transits}\label{sec:step}
Three \edit1{planets} listed in Table~\ref{tab:DMPP-1_RV_planets} (DMPP-1~b, b$^{\prime}$ and c$^{\prime}$) have periods which produce folded TESS light curves with incomplete phase coverage due to data gaps, or which could only exhibit a single transit.  However, the radii of these planets (implied by $M_{\mbox{\scriptsize P}} \sin i$) potentially allows single and partial transits to be detected from step-like changes in flux. Step detection was based on the difference between the mean fluxes in two equal-sized bins before and after a given time, with the significance of any change assessed in terms of the uncertainty ($\sigma_{\mbox{\scriptsize step}}$). We varied the bin width, from 4 to 216 cadences, i.e. 0.0056 to 0.30~d.  

No step-like features were detected in the TESS light curve at a level exceeding $5 \sigma_{\mbox{\scriptsize step}}$. The maximum value of $5 \sigma_{\mbox{\scriptsize step}}$ as a function of bin width is shown in Figure~\ref{fig:dip_step_limits}g--i (black line), with Table~\ref{tab:DMPP-1_RV_planets} giving the limit to any step change in flux (at a time-scale corresponding to maximum transit
duration) $(\Delta F/ F)_{\mbox{\scriptsize max}}$ as $5 \sigma_{\mbox{\scriptsize bin}}$.

As in Section~\ref{sec:fold} simulated light curves were used to determine the response to transits. Assuming the mass--radius relationship for pure water planets at $T=1000$~K given by \cite{Zeng19} the expected maximum ($b=0$) transit depths are  $\sim 600$~ppm, (solid cyan lines in Figures~\ref{fig:dip_step_limits}g--i), and would be detected if present. To be consistent with no detected signal, $b>0.97$ (dotted lines in Figure~\ref{fig:dip_step_limits}g--i; Table~\ref{tab:DMPP-1_RV_planets}). 

Alternatively, transits from DMPP-1~b, b$^{\prime}$ or c$^{\prime}$  may have been missed because their orbital periods exceed the continuous coverage the light curve. The coverage at periods of 18.57, 19.61 and 34.74 d is 94, 94, and 60 percent respectively. Thus there is a 6 percent chance that transits in DMPP-1~b and b$^{\prime}$ have been missed due to incomplete coverage, rising to a 40 percent chance for DMPP-1~c$^{\prime}$. 
\edit1{Further RV measurements would improve the precision of the orbital ephemerides and constrain the timings of the inferior conjunctions during the TESS coverage so these non-detections can be definitively interpreted.}

\subsection{Period searching using Transit Least Squares}
We find no evidence for transits at the orbital periods of DMPP-1~c, d, e, d$^{\prime}$, e$^{\prime}$, f$^{\prime}$, but further low-mass planets below the current RV detection threshold may exist. Hence, we searched for transits with $P<10.88$~d using the Transit Least Squares (TLS) method of \cite{2019A&A...623A..39H} implemented through the {\tt TLS} package. This fits a transit template (derived from \emph{Kepler} exoplanets) to provide improved sensitivity over the Box Least Squares (BLS) method \citep{2002A&A...391..369K}.  

\begin{figure}[htp]
    \centering
    \includegraphics[width=13cm]{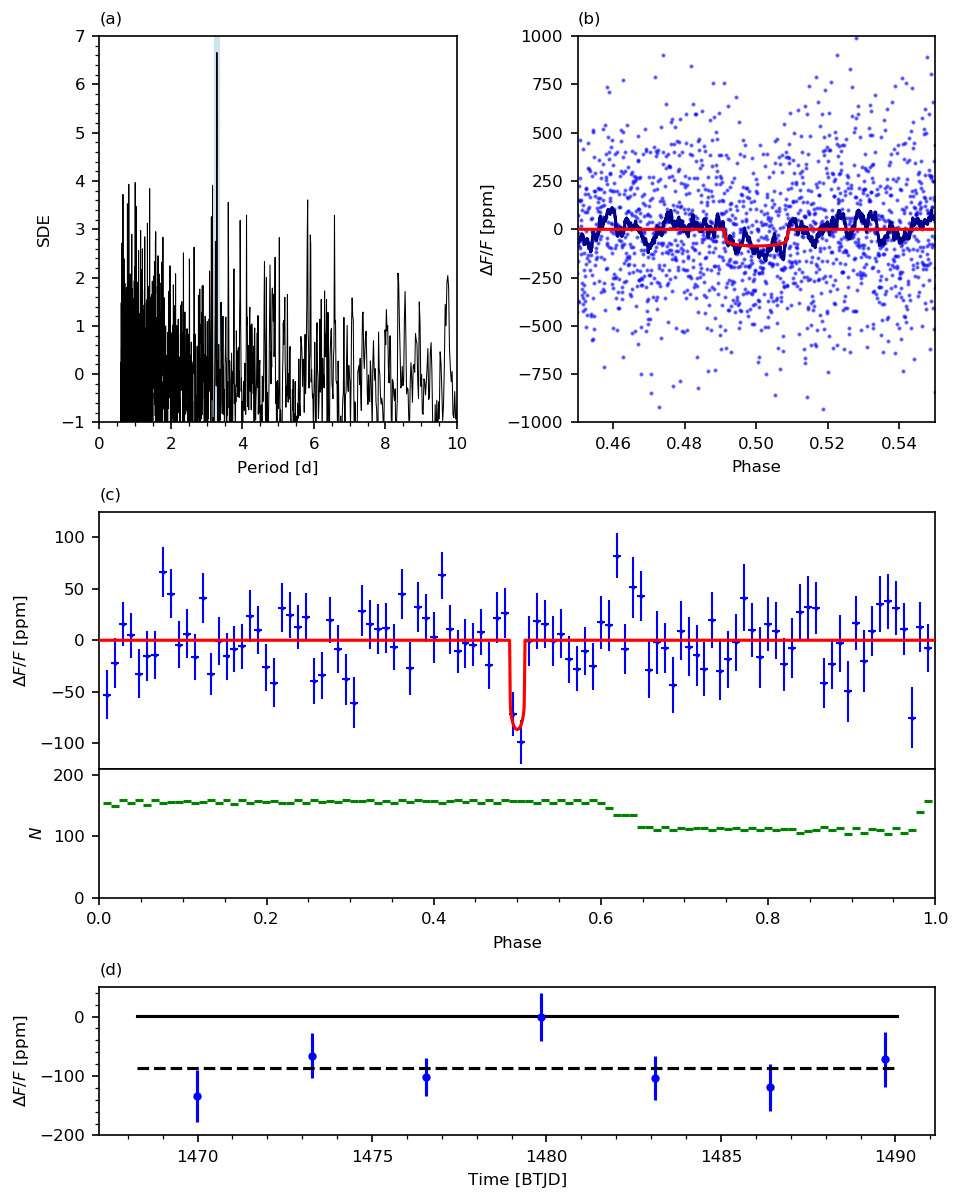}
    \caption{(a) SDE against trial period from a TLS search of the normalised TESS lightcurve of DMPP-1. (b) The lightcurve folded at the preferred solution of $P=3.2854$~d, with a moving mean ($N=41$~cadences) (dark blue trace) and the corresponding transit model (red). (c) The binned lightcurve folded at $P=3.2854$~d (blue) and transit model (red). The lower panel shows the number of cadences (green) in each phase bin. (d) The individual transits of the marginal signal with the depth of the preferred solution (87~ppm) shown by the dashed line. 
    }
    \label{fig:SDE}
\end{figure}

\begin{figure}[htp]
    \centering
    \includegraphics[width=13cm]{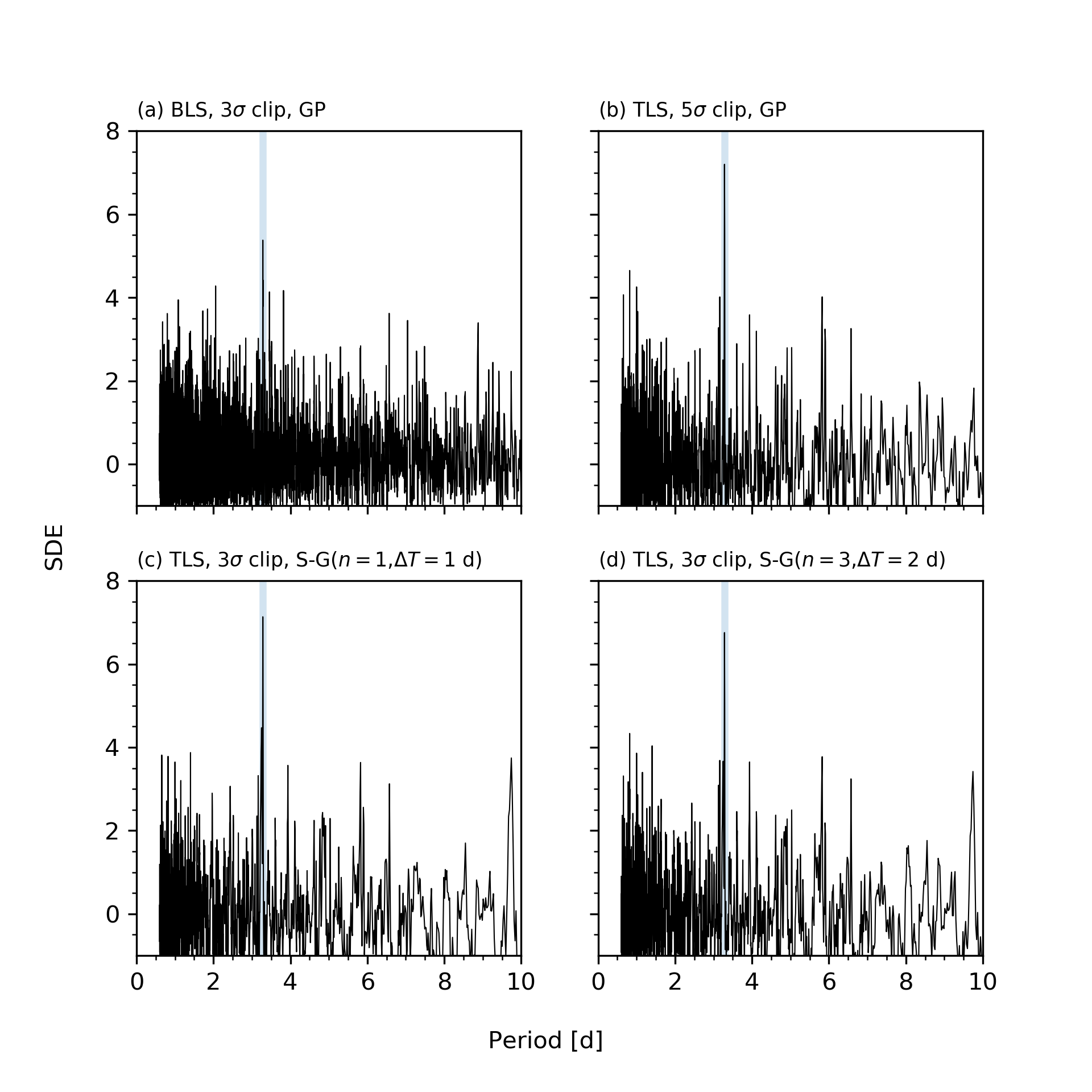}
    \caption{\edit1{A comparison of BLS and TLS periodograms of the normalised TESS lightcurve of DMPP-1 using different detrending methods. 
    (a) BLS -- $3\sigma$ clip and GP detrending, 
    (b) TLS -- $5\sigma$ clip and GP detrending,
    (c) TLS -- $3\sigma$ clip and S-G detrending ($n=1,\Delta T=1$\,d),
    (d) TLS -- $3\sigma$ clip and S-G detrending ($n=3,\Delta T=2$\,d)}
    }
    \label{fig:SDE_compare}
\end{figure}

The TLS period search on the normalised light curve of DMPP-1 yields SDE  
%\citep[signal detection efficiency,][]{2000ApJ...542..257A} 
\edit2{\citep[signal detection efficiency,][]{2002A&A...391..369K,2019A&A...623A..39H}} 
as a function of period as shown in Figure~\ref{fig:SDE}a. 
\edit1{The search covered $P= 0.60$~d \citep[the Keplerian period at the Roche limit  for a solar-type star, see][]{2014A&A...561A.138O} to $10.88$~d at with a frequency oversampling factor of 3 \citep{2019A&A...623A..39H}.}
While there are no peaks corresponding to the \cite{2019arXiv191210792S} planets, there is a signal at $P=3.2848\pm{0.0085}$~days with marginal significance, having SDE=6.66 and corresponding to a false-alarm probability (FAP) \edit1{of arising from white noise only} of 1.6\%. 
\edit1{A BLS search \edit2{\citep[with SDE calculated as described in][]{2002A&A...391..369K}} on the same data retrieves a maximum signal at $P=3.2847$\,d but with a lower value SDE=5.2 (Figure~\ref{fig:SDE_compare}a) as expected \citep[see][]{2019A&A...623A..39H}.}

Parameters of the best-fitting transit model (implemented through {\tt batman}) were found by maximum likelihood estimation, confirming the above period. Values and 16--84\% credibility regions 
%(16--84 percent) for model parameters 
were found by sampling the posterior probability distribution using MCMC \citep[implemented with {\tt emcee},][]{2013PASP..125..306F} assuming Gaussian priors on $P$ (from the TLS result) and stellar density \citep[using the stellar mass and radius values of][]{2019arXiv191210792S}; and uniform priors for
other parameters.
%were assumed to have uniform priors). 
Estimating parameter values from the medians of the posterior distributions gives a preferred solution of  $P=3.2854^{+0.0032}_{-0.0025}$~d, 
%and 
$R_{\mbox{\scriptsize{P}}}=0.00100^{+0.0021}_{-0.0023} R_{\ast}$ $=1.38^{+0.29}_{-0.32} R_{\oplus}$ and $\delta=87^{+25}_{-30}$~ppm, while the stellar density is $\rho_{\ast}=0.61\pm{0.03} \rho_{\odot}$  (see 
%are 
%summarised in 
Table~\ref{tab:DMPP-1_RV_planets}c
for other parameters).

{}

The folded light curve is shown both as cadences and running mean ($N=41$) (Figure~\ref{fig:SDE}b) and binned
(Figure~\ref{fig:SDE}c) along with the best-fitting transit model. 
%It is notable that t
The putative transit does not correspond to changes in the number of cadences per bin (Figure~\ref{fig:SDE}c) as might be expected from ineffective detrending near the start or end of the light curve.  
%Not surprisingly, 
The seven individual transits 
%(of which there are seven) 
have signal-to-noise ratios between 0.03 and 3.06 (Figure~\ref{fig:SDE}d).

{}
Since this is a marginal signal, we investigated the sensitivity of the detection to the pre-processing carried out on the light curve. Firstly, we varied the level at which sigma-clipping is applied: at $5 \sigma$, 7 data points are excluded (as compared to 69 above), and the 3.2848~d signal has an enhanced SDE of 7.20 \edit1{(Figure~\ref{fig:SDE_compare}b)}. The likely cause is that the additional 62 data points raise the out-of-transit level, thus giving a greater significance to the dip. We conclude that a $3 \sigma$ clipping level is a conservative but appropriate approach.

We investigated the sensitivity to the detrending method by using two other commonly-used methods: Savitzky-Golay (S-G) and a sliding window Tukey biweight estimator.  S-G filters \citep[see e.g.][]{2011ApJS..197....6G} are specified by the order ($n_{\mbox{\scriptsize S-G}}$) of the detrending polynomial, and a window-width, 
%(specified by number of cadences, but for convenience 
specified here as duration in the absence of data gaps, $\Delta T_{\mbox{\scriptsize S-G}}$
%) 
. We chose
%adopted 
ranges: $n_{\mbox{\scriptsize S-G}}=[1, 2, 3]$ and $\Delta T_{\mbox{\scriptsize S-G}}=[0.25, 0.5, 1.0, 2.0, 3.0]$~d %(chosen 
to span different timescales of variability, up to a significant fraction of the duration of contiguous data.
%).  
9 of the 15 combinations
%9 
returned a maximum SDE at a period of 3.2848~d. 
%The range of t
These SDE values ranged
%is 
from 7.13 to 4.96 with corresponding FAPs from 0.76$\%$
to $ > 10\%$. 
\edit1{Two of these S-G filters ($n=1$,$\Delta T$=1\,d and $n=3$,$\Delta T$=2\,d, Figure~\ref{fig:SDE_compare}c, d respectively) returned SDE values higher than that obtained with the GP filter.} 
%above 10~percent. 
None of the light curves with $\Delta T_{\mbox{\scriptsize S-G}}=0.25$~d recovered this signal, probably because the dip timescale is similar. 

A filter based on the Tukey biweight estimator \citep{1977dars.book.....M} was implemented through the {\tt W{\={o}}tan} package \citep{2019AJ....158..143H}. 
%Again, a range of 
Window widths 
$\Delta T_{\mbox{\scriptsize biw}}=[0.25, 0.5, 1.0, 2.0, 3.0]$~d were adopted, with edge truncation of half the window width, and tuning parameter $c=6$, 
%\citep[as being 
appropriate for transit light curves \citep{2019AJ....158..143H}. 
%In this case, 
Only 
%one filter (
$\Delta T_{\mbox{\scriptsize biw}}=1.0$~d
%) 
returned the $P=3.2848$~d signal, and this with SDE=4.71 and FAP $>$~10 $\%$.

\edit1{It should be stressed that} GP modelling is our preferred detrending method because the variability timescale is determined from the data (rather than arbitrary choices for $\Delta T_{\mbox{\scriptsize S-G}}$ or $\Delta T_{\mbox{\scriptsize biw}}$).
Nonetheless, the sensitivity to the detrending methods 
implies 
%means that 
the 3.2854 d signal 
%at should be considered 
is a marginal detection.
%and that f
Further data are needed to confirm
%the existence of a transiting body 
transits at this period. 

\subsection{Limits and significance of modulation}
\begin{figure}
    \centering
    \includegraphics[width=15cm]{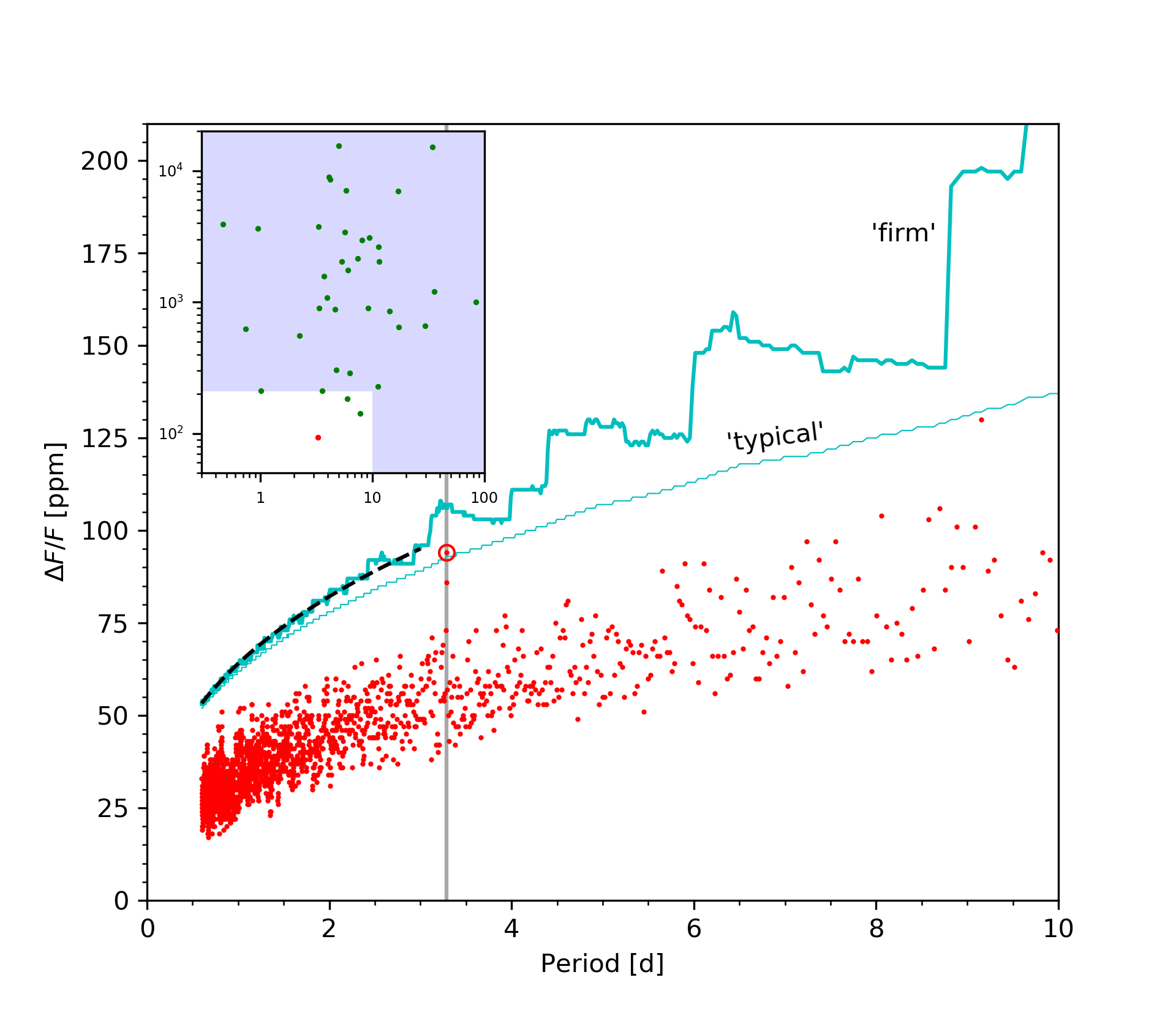}
    \caption{\edit1{Measurements and limits of the amplitude of modulation in the TESS light curve of DMPP-1. Actual modulations (red dots) are the maximum dip-like changes in a single phase bin (corresponding to $0.4 T_{\mbox{\scriptsize{dur, max}}}$) at a given period. The thick cyan line shows the firm upper limit to modulation ($A_{5 \sigma,\mbox{\scriptsize{max}}}$) while the thin cyan line shows the typical level of modulation ($A_{5 \sigma,\mbox{\scriptsize{mean}}}$) that is potentially detectable. 
    The dashed black line at $P<3$~d shows an approximate analytical form to the firm amplitude limit. The red circle shows the modulation at $P=3.2854$~d. (Note that the modulation at 9.16\,d that is near the `typical' detection curve has a significance of 3.3$\sigma_{\mbox{\scriptsize{bin}}}$)
    The inset shows the context of the $P=3.2854$~d signal (red dot) in relation to all TESS exoplanets (green dots) with amplitudes calculated as $(R_{\rm P}/R_{\ast})^2$. }
    %The green stars on the main figure show the short-period TESS planets HD21749c \citep[$P=7.78$~d,][]{2019ApJ...875L...7D} and HR858c \citep[$P=5.97$~d,][]{2019ApJ...881L..19V}. 
    }
    \label{fig:modulation_limits}
\end{figure}

%In the absence of a strong transit signal it is useful to quantify the limits to any signal that may be present. 
\edit1{The period folding method of Section~\ref{sec:fold} was used to find transit detection limits on the same period grid as our TLS search.
Noting that such limits are a compromise between sensitivity and measurement timescale, we adopted a bin-width of $0.4 T_{\mbox{\scriptsize{dur, max}}}$ (i.e. sensitive to transits with $b<0.917$) since this is expected to show the modulation due to the 3.2854\,d signal and allow its significance to be estimated independently of the SDE value returned by TLS. 
%Figure~\ref{fig:modulation_limits} shows two limits to modulation. 
In general, at a given folding period, the value of $\sigma_{\mbox{\scriptsize{bin}}}$ varies with phase (c.f. Fig.~\ref{fig:SDE}c). 
Consequently we show two quantities in Figure~\ref{fig:modulation_limits}:
$A_{5 \sigma,\mbox{\scriptsize{max}}}$ (thick cyan line) and $A_{5 \sigma,\mbox{\scriptsize{mean}}}$ (thin cyan line) defined as 5 times the maximum and mean values of $\sigma_{\mbox{\scriptsize{bin}}}$ respectively.  
%At short periods, $A_{5 \sigma,\mbox{\scriptsize{max}}}$ and $A_{5 \sigma,\mbox{\scriptsize{mean}}}$  have similar behaviours. 
In the absence of a detected signal, $A_{5 \sigma,\mbox{\scriptsize{max}}}$ provides the firm upper limit to any modulation that may be present, and at short periods ($P<3$~days) it is well-represented by 
\begin{equation}
    A_{5 \sigma,\mbox{\scriptsize{max}}} = 64 ( P/ \mbox{days} )^{0.36} \mbox{ppm}
    \label{eqn:A5sigma_limit}
\end{equation}
 (dashed line in Figure~\ref{fig:modulation_limits}). At longer periods $A_{5 \sigma,\mbox{\scriptsize{max}}}$ exhibits step-like behaviour with increasing $P$, arising from large changes in the minimum number of cadences per phase bin.  $A_{5 \sigma,\mbox{\scriptsize{mean}}}$ indicates the typically detectable depth threshold. }
\edit1{ 
Figure~\ref{fig:modulation_limits} also shows the maximum detected modulation (red dots) of all transit-like (i.e. dip) signals. Also shown (red circle) is the result of folding at 3.2854\,d: this yields $\Delta F / F = 94$\,ppm and $5.4\sigma_{\mbox{\scriptsize{bin}}}$ significance ($1.01 \times A_{5 \sigma,\mbox{\scriptsize{mean}}}$). The FAP for a $5\sigma_{\mbox{\scriptsize{bin}}}$ variation in this search for DMPP-1 is calculated by noting that there are 70217 independent combinations of period and phase bin \citep[the number of independent periods was found using Equation 6 of][]{2014A&A...561A.138O}. Multiplying this by the one-sided probability of a 5$\sigma$ deviation ($2.87\times 10^{-7}$) gives FAP = 2.0\%. Since the signal exceeds $5\sigma_{\mbox{\scriptsize{bin}}}$, FAP $<$ 2.0\% and is consistent with the value found by TLS.}
%Formally, the corresponding FAP at $5.4 \sigma_{\mbox{\scriptsize{bin}}}$ is 0.2\% although given the sensitivity of the signal to detrending it is preferable to state that the FAP$<2$\% noting that this is consistent with the value returned by TLS.

\section{Discussion}\label{disc}
% MHJ Attempt to structure argument
% Key question - can we place any constraints on the source of the circumstellar material in DMPP-1?
% 
% Firstly, set the context in terms of insolation
% Then innermost RV planets may seem obvious source of material - but it might not be this simple - 
% Possibilities:
% 1) Innermost RV planet is source of shroud 
%   No transits detected implies: 
%   EITHER 
%       b>b limit => gas cloud is extended
%   OR 
%       delta<detection limit gas cloud has lower extent
%       and planet is dense
% 2) 3.28 d planet is real and is source of shroud
%       REAL 
%           Orbital stability implies M<3.5 ME
%           Require near circular orbits
%           Lack of RV signal implies small M
%       SOURCE OF SHROUD 
%           For this planet to be ablating. M<<M(interior)
%           Some hint that transit level is variable.
%           gas cloud is not extended
%           maybe a CDE
% 3) An undetected planet interior to innermost planet is ablating
%   NO TRANSIT implies
%   EITHER 
%       (rp/r*)^2 below detection limit and cloud is thin
%   OR 
%       No transits occur and cloud is thick.
% 

\subsection{No transits of RV planets}
% No transits detected except a marginal signal that is not confirmed by RV measurements
The TESS photometry does not reveal any transits corresponding to the planets reported by \cite{2019arXiv191210792S}. 
If these planets are of Earth-like rocky composition,  or have compositions similar to Solar System ice giants, their radii imply transit depths for edge-on orbits which  exceed our detection thresholds. % MHJ rewritten to address ref's comment, but not happy with this sentence
\edit1{We conclude that either the planets are not in orbits that transit the host star, or that no transit of (Neptune-like) planets happened to occur during the TESS observation, or that (super-Earth) planets are smaller than models based on Solar System planets \citep{Zeng19} would predict.}

% MAYBE INCLUDE THIS?
\edit1{The extent of the circumstellar gas cloud can be estimated assuming that the lack of observed transits is due to orbital inclination. As noted by \cite{2019arXiv191210792S}, the cloud could originate from any of the super-Earth planets (-1c, d, e, or d$'$, e$'$, f$'$). Adopting a simple geometric model in which the gas cloud extends a height $h$ vertically from the orbital plane, then the requirement that the cloud completely covers the stellar photosphere but that the planet does not transit implies $h\geq 2R_{\ast} / [1-(R_{\ast}/a)^{2})]^{1/2}$. Assuming azimuthal symmetry, the solid angle subtended by the cloud from the centre of the star is $\Omega_{\mbox{\scriptsize{c}}} \geq 4 \pi (2/[3+(a/R_{\ast})^{2}]^{1/2})$. The fraction of the sky covered by the cloud $>27\%$ if it originates from DMPP-1\,d with $P\approx 3$~d ($>16\%$ for DMPP-1\,c with $P\approx 6.5$~d).} 
%
% We conclude that either these planets were not in orbits which transited the host star at the time of the TESS observations, or the planets are smaller than models based on Solar System planets \citep{Zeng19} would predict.

\subsection{Insolation of DMPP-1 planets and known CDEs}
\label{sec:insolation}
Supplementary Figure 4b of \cite{2019arXiv191210792S} 
%makes it clear 
establishes that the DMPP-1 planets are among the most irradiated known low mass planets orbiting bright, nearby stars. They fall along the lower boundary of the Neptune desert.
The insolation values, $S$, of DMPP-1\,b,  c,  d  and e are tabulated in Table~\ref{tab:DMPP-1_RV_planets}. 
%The hottest of these planets, DMPP-1\,d, receives almost 1200 times the flux Earth receives from the Sun. Mercury has an insolation of about $7\, {\rm S_{\oplus}}$, and has an enhanced iron content of $70\,\%$ (by mass). It is unclear if these characteristics are directly related, as the erosion of Mercury's mantle could have been caused by a single high energy impact \citep{Benz2008,Marcus2010}. Recently, a number of ultra-short period exoplanets including KOI\,1843.03, K2 137\,b, K2 229\,b, and K2 106\,b have been inferred to have similar enrichments of iron \citep{PriceRogers2019}. One possibility for this iron enrichment is vaporisation of the rocky surface due to the intense insolation. \cite{Kite2016} model this interior-surface-atmosphere-wind mass exchange in detail. 
%DMPP-1 is much more luminous than the typical Kepler and K2 host star; the planets orbiting DMPP-1 are expected to have temperatures at the sub-stellar point similar to those of planets in sub-day orbits around K dwarf stars.
 DMPP-1\,d's
 %in a 2.88~d orbit receives comparable 
 insolation is similar 
 %flux, 1177~$S_{\oplus}$ [{\bf need to propagate errors}],  
 to that of Kepler-1520\,b 
 %in a 15.7~hour orbit around a K dwarf 
 %\citep[1134.386$^{+56.848}_{-53.974}S_{\oplus}$,][]{2018ApJ...866...99B}. 
 \citep[1134$^{+57}_{-54}S_{\oplus}$,][]{2018ApJ...866...99B}. 
 % Kepler-1520\,b is the most dramatic known example of mass loss from a rocky planet. 
 The other two known CDEs: K2-22b \citep{2015ApJ...812..112S} and KOI-2700b \citep{2014ApJ...784...40R} are less irradiated with insolation fluxes of 
 %$737.9^{+242.8}_{-196.7} S_{\oplus}$ \citep{2017AJ....154..207D} 
 $738^{+243}_{-197} S_{\oplus}$ \citep{2017AJ....154..207D} 
 and $801^{+32}_{-31} S_{\oplus}$ \citep{2018ApJ...866...99B} respectively. 
 Coupled with the circumstellar gas which apparently surrounds DMPP-1, this constitutes grounds to suspect that the planets orbiting DMPP-1 may have been significantly eroded as a consequence of proximity to their luminous F8V host star. The DMPP-1 planets are probably not pure iron, but they may not be very much bigger than this limiting model. 
 Table~\ref{tab:DMPP-1_RV_planets} and Figure~\ref{fig:modulation_limits} show that planets DMPP-1~d, e, e$^{\prime}$ and f$^{\prime}$ are either undetectable or close to the detection threshold.
 We need to drive down the detection thresholds with more precise photometry, e.g. with CHEOPS observations. 
 %than the single sector of TESS data available so far. CHEOPS observations would be ideal for searching for transits of these planets.
 
 \subsection{Marginal Detection of a P=3.2854~day CDE?}
 We 
 %have made a 
 marginally detect
 a transit signal with $P=3.2854$~d in the TESS data. 
 %While we cannot be confident that this is a genuine planetary signal, we can usefully explore the implications of a putative planet with this period (hereafter `Planet 5') being present.  
 %The FAP of this signal is sensitively dependent on the choices made for removing long-term trends in the data. 
 If this were due to an Earth-like rocky planet, it would correspond to 
 %or a planet with enhanced iron content, it would have been apparent in the extant RV data. Using the Earth-like rocky planet models of \cite{Zeng19}, this TESS signal corresponds to 
 a planet of mass $3.2^{+3.6}_{-2.0} \, {\rm{M_{\oplus}}}$. 
 %A re-inspection of 
 The RV data of 
 %presented by 
 \cite{2019arXiv191210792S} shows no evidence for modulation at this period.
 %, and the detection cannot be confirmed.
 
%Regarding the stability of the orbit, 
%Since DMPP-1d and the putative 3.28\,d Planet~5 are in similar orbits, 
We carried out orbital integrations, using {\tt IAS15} within {\tt REBOUND} \citep{2015MNRAS.446.1424R, 2012A&A...537A.128R},
for the DMPP-1 system of 4 planets,
adding the putative 3.2854\,d Planet~5.
%with  variable mass.  
%We used the {\tt IAS15} integrator \citep{2015MNRAS.446.1424R} within the N-body orbital integrator, {\tt REBOUND} \citep{2012A&A...537A.128R}. 
Assuming circular orbits ($e = 0$), and adopting a range of masses for Planet~5, we find that all configurations remain stable for $10^5$ yrs for masses $\leq\ 3.5\, {\rm M_{\oplus}}$ 

Since \cite{2019arXiv191210792S} found upper limits to the eccentricities of $e=0.083,\ 0.057,\ 0.07,\ 0.07$ for DMPP-1\,b, c, d, e, we also carried out eccentric orbital integrations at these limits including Planet~5 (with $e=0.07$). As expected, the system is unstable for all masses of Planet~5 since it crosses the orbit of DMPP-1\,d. Considering the 0.006 AU Hill sphere of DMPP-1\,d and assuming a negligible mass for Planet~5 indicates an upper eccentricity limit of $e = 0.04$ for both planets. The true value will be smaller owing to the influence of DMPP-1\,b, c and e. Hence if the $3.2854$ d planet exists, the orbits of all DMPP-1 planets must be very close to circular, with somewhat lower eccentricities than the upper limits reported by \cite{2019arXiv191210792S}.

 A planet in a circular orbit around DMPP-1 with P=3.2854~d would suffer an insolation of $990 \, {\rm S_{\oplus}}$, typical for a
 %which is in the middle of the range of fluxes experienced by the three known 
 CDE.  The marginal evidence for variable transit depth in Fig.~\ref{fig:SDE}d could indicate that the transits are akin to those of Kepler-1520\,b. If this were the case, we would expect a tiny reflex RV modulation.

In the case of Kepler-1520\,b, transits are due to a dust cloud
\citep{2012ApJ...752....1R,2012A&A...545L...5B,2015ApJ...800L..21B} rather than the planet itself.  If the $P=3.2854$~d signal arises in an analogous way, then assuming grain radius $s$, dust density $\rho$, \edit1{and dust sublimation rate $\tau_{\mbox{\scriptsize{dust}}}$,} 
we estimate the dust mass as
%an estimate of the dust mass can be made by assuming that grains have a characteristic radius $s$  \citep[$\sim 1 \mu$m,][]{2015ApJ...800L..21B} and that extinction is approximated by the geometrical limit. Ignoring limb-darkening effects, a transit of depth $\delta $ implies a dust mass
%\begin{equation} \label{eqn:Mdust}
%    M_{\mbox{\scriptsize{dust}}} \approx \frac{4\pi}{3}  R^{2}_{\ast}s \rho \delta
%\end{equation}
%where $\rho$ is the grain density. Applied to this signal, 
\begin{equation}
    M_{\mbox{\scriptsize{dust}}} \approx 9 \times 10^{14} \left(\frac{\rho}{\mbox{3 g cm$^{-3}$}}\right) \left(\frac{s}{\mbox{ 1$\mu$m}} \right) \mbox{g},
\end{equation}
and 
%Dust in Kepler-1520b is expected to sublimate on timescale $\tau_{\mbox{\scriptsize{dust}}}\sim 3 \times 10^{4}$~s \citep{2012ApJ...752....1R}, so if a similar process operates in DMPP-1, 
the rate of dust production as
\begin{equation}
    \dot{M}_{\mbox{\scriptsize{dust}}} \sim 3 \times 10^{10} \left(\frac{\rho}{\mbox{3 g cm$^{-3}$}}\right) \left(\frac{s}{\mbox{ 1$\mu$m}} \right) \left(\frac{\tau_{\mbox{\scriptsize{dust}}}}{3 \times 10^{4}\mbox{s}} \right)^{-1}
    \mbox{g s$^{-1}$,}
\end{equation}
%Thus, dust production by a planet with $P=3.28$~d could produce the marginal transit signal if the dust production rate is 
about one third of %that implied for 
Kepler-1520\,b's value, $ \dot{M}_{\mbox{\scriptsize{dust}}} \sim 10^{11}\mbox{g s}^{-1}$
\citep{2012ApJ...752....1R}.

\subsection{Metal-rich gas from ablating planets}
\label{CDEgas}
Metal-rich gas must co-exist with the dusty effluents from CDEs  \citep{2012ApJ...752....1R, RH19}. A search for traces of this circumstellar gas in the Na\,D and Ca II infrared triplet lines in spectra of K2-22 yielded null results and a range of upper limits \citep{RH19}. 
%K2-22 is the brightest of the three known CDEs, and the most favourable of them for the search for gas. However, the approach taken was to perform transmission spectroscopy through the differencing of in transit and out of transit spectra. If the gas is azimuthally near-homogeneous, rather than being strongly concentrated around the azimuth of the planet, this is not the ideal strategy.  \cite{Haswell2012} discovered that the chromospheric Ca\,II\, h\&k emission in WASP-12 is  completely absorbed at all observed orbital phases, suggesting that gas extends all the way around the star, rather than being localised at the azimuth of the ablating planet. The success of the DMPP target selection, based on Ca\,II\,H\&K measurements made at arbitary times \citep{2019NatAs.tmp....2H}, further suggests that the conventional transmission spectroscopy approach is not the best way to detect this gas. DMPP's target selection criteria rely on picking out anomalous absorption of the intrinsic stellar spectrum. An approach similar to this might reveal the metal-rich gas component of the planetary mass loss from K2-22\,b.
DMPP-1 is already known to exhibit circumstellar absorption \citep{2019arXiv191210792S,2019NatAs.tmp....2H}, so if the marginal transit detection reported herein does prove to be a CDE, this system will be the best yet known to trace the gas loss and sublimation processes discussed in \cite{RH19}. DMPP-1 is almost 8 magnitudes brighter than K2-22 in the $V$ band. Thus this system offers exciting scope for high temporal and spectral resolution studies. 

\subsection{Hot rocky planet compositions}
The atmospheric compositions of hot rocky planets have been modelled by
%Work modelling the composition of the atmospheres of hot rocky planets began over a decade ago: 
\cite{SchFeg09,Miguel11,Schaefer12,Lupu14,Ito15}. 
%considered a molten magma ocean with a silicate atmosphere resulting from the gas-melt equilibrium. 
%\cite{Miguel11} 
%extended this work, with models for surface temperatures ranging from 1000-3500\,K, planet masses from $1-10\,\rm{M_{\oplus}}$, and magma with bulk silicate Earth (BSE) composition or komatiite. Komatiite has low Si, K, and Al abundance, elevated Mg abundance, and a melting point $ \sim 1900\,K$. Terrestrial komatiite samples have ages $> 2.5 \rm{Gyr}$. The abundance differences between komatiite and present-day lava are attributed to crystallisation from the hotter Archaean mantle melts. 
%
%The properties and detectability of volatile-rich mineral atmospheres  \citep{Schaefer12,Lupu14} and volatile-free mineral atmospheres \citep{Ito15} have been considered. \cite{Ito15} compile spectra line data for the major constituents of volatile-free mineral atmospheres and calculate the emission spectra for a super-Earth with $g = 25\,{\rm m\,s^{-2}}$ and sub-stellar point equilibrium temperatures ranging from 1800\,K to 3000\,K. Na is generally the most common constituent of the atmosphere, with the molar fraction of SiO rising with temperature to dominate at the highest temperatures considered for all four magma compositions considered. 
The atmospheric absorption in the UV and IR is dominated by SiO, and the short wavelength absorption causes a temperature inversion, promoting the formation of strong spectral features. 
%The thermal inversion allows the detection  of the 10\,$\mu$m SiO features from hot nearby super-Earths via JWST secondary eclipse data
%with JWST out to a distance of 30\,pc of  from the atmosphere of a ${R= 2 \rm{ R_{\oplus}}, T_{eq}= 2300\,{\rm K}}$ super-Earth orbiting a solar analogue. 
%\cite{Ito15} note that transmission spectroscopy is more challenging because of the small scale-height and hence tiny absorbing annulus of the mineral atmosphere of a transiting terrestrial planet.
%
%The CDEs are much more promising targets for observational compositional studies as the ablating material is dispersed over scale-heights comparable to the stellar radius. The signal to noise ratio for transmission spectroscopy is proportional to the scale-height. As alluded to in Section~\ref{CDEgas}, the Ca\,II and Mg\,II from the DMPP planets and from WASP-12\,b appears to be well-mixed azimuthally: in this case transmission spectroscopy will not work. With SiO, however, the UV flux from the host star will cause photo-dissociation, so SiO absorption will be localised around the planet.
\cite{Kite2016} model the processes governing the gas and dust escape from a hot rocky planet. They find two cases: magma-ocean-dominated and silicate-atmosphere dominated. In the latter case, which applies to the hottest planets with equilibrium temperature,
$T_{\rm eq} >  2400 \, {\rm{K}}$, wind transport dominates and the magma pools are compositionally patchy. The chemical abundance mix in the mass loss can consequently be variable. The outflow from the putative CDE causing our marginally detected transit signal could potentially allow us to directly probe this compositional variability.

Surface compositional heterogeneity is not fully understood even for the Earth \citep{Rizo16}, and may provide traces of the primary accretion phase of planet formation. The potential to observe the surface composition variations of exoplanets offers a valuable counterpoint to the Solar System based picture of planetary system formation. 

\section{Summary and Conclusions}\label{conc}
\begin{enumerate}
    \item 
DMPP-1 is a bright, nearby star, hosting a compact multi-planet system. We expect the DMPP-1 planetary orbits to be aligned approximately edge-on as ablated planetary gas fills our line of sight to the chromospherically active regions of the star.

\item
We analysed the TESS data for DMPP-1, performing
a sensitive search for transits of the RV planets reported by \cite{2019arXiv191210792S}. We find null results, suggesting that perhaps these planets are denser than conventional Earth-like rocky planets. 
%Other hot, short period exoplanets, and even Mercury in the Solar System, have much higher iron fractions than Earth.
%
%\item
Alternatively, the 
%explanation for our null results is that 
DMPP-1 RV planets are not exactly 
%is not in an 
edge-on.
%orientation. In this case we can place lower limits on the vertical extent above and below the of the ablated gas. {\bf put in limits}

\item
We marginally detect (with FAP of $1.6$\%) a 87$^{+25}_{-30}$\,ppm transit with a period of $P=3.2854^{+0.0032}_{-0.0025}$\,d. One of the 7 transits has a depth consistent with zero, reminiscent of the variable transit depths of the known CDEs.

\item
The $P=3.2854^{+0.0032}_{-0.0025}$\,d period corresponds to 
%a circular Keplerian orbit with 
insolation $S = 990 \rm{S_{\oplus}}$, typical of CDEs. 
%This is in the middle of the range of fluxes experienced by the three known CDEs. This transit signal could self-consistently be attributed to 
A CDE would be 
%with dusty effluents and mass 
well below the RV detection threshold.
%of \cite{2019arXiv191210792S}.

\item
The detection of 
%significance with which we detect 
the $P=3.2854^{+0.0032}_{-0.0025}$\,d signal is sensitive to
%ly dependent on 
the detrending approach adopted.
%used to remove low amplitude stellar astrophysical variability. 
We therefore consider our detection marginal. 
Small planets orbiting solar type stars are challenging to find in a single sector of TESS data.
%requires care.
%will benefit from careful examination of the effects of the various   detrending algorithms. 
Further sectors of TESS data and targeted photometry with CHEOPS will be helpful in the assessment of marginal signals.

\item
JWST will be very sensitive to SiO gas which is expected in material ablated from hot rocky planets including  CDEs. DMPP-1 \edit1{could be} an excellent candidate for IR spectroscopy with JWST to examine hot rocky planet compositions. 
\end{enumerate}
% If the source of the gas cloud is the innermost of these planets ($P\sim 3$~days), then the cloud covers at least 25~percent of the sky as viewed from the stellar centre. Alternatively, the gas may originate from an undetected sub-Earth sized transiting planet, in which case, the vertical extent of the cloud may be about half that of the non-transiting case, and the solid angle covered would be correspondingly smaller. In this latter scenario, the lack of transits implies that any dust production must be at much lower rate than in Kepler~1520b. 

\section*{Acknowledgements}
CAH and JRB are supported by STFC Grants  ST/P000584/1 and ST/T000295/1.
DS was supported by an STFC studentship.

\end{document}